\documentclass{article}
\begin{document}
\title{General relativistic spinning fluids with a modified projection tensor}
\author{Morteza Mohseni\thanks{E-mail address:m-mohseni@pnu.ac.ir}\\
\small Physics Department, Payame Noor University, 19395-4697
Tehran, Iran}

\maketitle
\begin{abstract}
An energy-momentum tensor for general relativistic spinning fluids compatible with Tulczyjew-type supplementary condition is derived from the 
variation of a general Lagrangian with unspecified explicit form. This tensor is the sum of a term containing the Belinfante-Rosenfeld tensor 
and a modified perfect-fluid energy-momentum tensor in which the four-velocity is replaced by a unit four-vector in the direction of fluid 
momentum. The equations of motion are obtained and it is shown that they admit a Friedmann-Robertson-Walker space-time as a solution.

keywords: Spinning fluids, Spinning particles

PACS: 04.20.-q, 04.20.Fy
\end{abstract}
\section{Introduction}
According to the general theory of relativity, free particles move along trajectories which are geodesics of the background space-time. 
Kinematically, the trajectory of each particle can be characterised by the particle four-velocity which is a time-like vector tangent to its 
path. This tangent also shows the direction of the particle momentum. On the other hand the dynamics of a general relativistic perfect fluid can 
be described by its energy-momentum tensor. This energy-momentum tensor is expressed in terms of the fluid energy density and pressure and a 
projection tensor which is itself constructed out of the four-velocity of the flow-lines and the space-time metric.

When the particle has an internal classical spin, it no longer moves along a geodesic path in general due to a coupling between the particle 
spin and the space-time curvature. The dynamics of a spinning particle is described by the Mathisson-Papapetrou-Dixon (MPD) equations 
\cite{dix}, according to which the particle four-momentum is time-like but not co-linear with its four-velocity. Thus in treating the dynamics 
of a spinning particle, we deal with two distinct time-like four-vectors corresponding to the velocity and momentum four-vectors. These two 
preferred time-like vectors have been used to construct the two most favoured supplementary conditions (needed to render the MPD equations 
complete), the so-called Pirani condition and the Tulczyjew condition, see e.g. \cite{dix}. In the former the particle spin tensor has no 
electric components in the instantaneous zero-velocity frame while in the latter it has no electric components in the instantaneous 
zero-momentum frame.     

Several extensions of perfect fluid to spinning fluids, i.e. perfect fluids whose particles have an internal but classical spin, have been made 
in the literature basically by adding terms containing spin to the perfect-fluid energy-momentum tensor \cite{ray,sma,obu,bai}. 
The interrelations between the dynamics of spinning particles and the dynamics of spinning fluids have been studied in \cite{kop} in the 
framework of Einstein-Cartan theory. In these studies always a Pirani-type supplementary condition has been used. The question then arises if a 
formulation of spinning fluids consistent with Tulczyjew condition is possible. The aim of the present work is to address this question. We show 
that such a formulation is possible provided the projection tensor which constitutes the energy-momentum tensor, is constructed out of a time-
like four-vector in the direction of the momentum, rather than the velocity four-vector. 

To construct such a spinning fluid model, we deploy the machinery developed in ref. \cite{bai} in which a spinning fluid energy-momentum has 
been derived from a general Lagrangian whose explicit form is not specified. We start with such a Lagrangian and end up with an energy-momentum 
tensor whose (explicitly) spin-dependent terms are similar to the corresponding terms in the above mentioned reference but in the remaining part 
the usual perfect-fluid projection tensor is replaced with a modified one. In this modified projection tensor the four-velocity is replaced with 
a unit four-vector in the direction of the four-momentum.

In the following sections we first briefly review the MPD equations describing the motion of spinning particles in curved space-times. Then we 
follow a customised version of the procedure developed in \cite{bai} to obtain the energy-momentum tensor. The equations of motion can then be 
obtained by imposing the requirements of symmetry and conservation on the energy-momentum tensor. We use the resulting equations of motion to 
study the dynamics of the spinning fluid in a flat Friedman-Robertson-Walker universe generated by the fluid. We show that these equations admit 
a solution of this type. A discussion of the results concludes the main sections. Some supplementary calculations are collected in an appendix.

\section{The MPD Equations}
The motion of a spinning particle moving in a curved space-time is described by the so-called MPD equations. Neglecting the particle multi-poles 
higher than dipole, these are given by \cite{dix}
\begin{eqnarray}
{\dot s^{\mu\nu}}&=&p^\mu v^\nu-p^\nu v^\mu,\label{eq3a}\\
{\dot p^\mu}&=&-\frac{1}{2}{R^\mu}_{\nu\alpha\beta}v^\nu
s^{\alpha\beta},\label{eq3b}
\end{eqnarray}
where $s^{\mu\nu}$ are the particle spin tensor, $p^\mu$ are the particle momentum, $R_{\mu\nu\kappa\lambda}$ represent the curvature tensor, 
over-dots mean covariant differentiation $v^\alpha\nabla_\alpha$, and $v^\mu$ are the particle velocity. These equations supplemented by
the so-called Tulczyjew condition
\begin{equation}\label{eq4}
p_\mu s^{\mu\nu}=0
\end{equation}
result in
\begin{eqnarray*}
p_\mu p^\mu&=&\mbox{const.}=-m^2,\\
s_{\mu\nu}s^{\mu\nu}&=&\mbox{const.}=2s^2
\end{eqnarray*}
where $m, s$ are the particles mass and spin respectively. Hence, defining $$\chi^\mu=\frac{1}{m}p^\mu$$ the normalization 
$$\chi_\mu\chi^\mu=-1$$ is
guaranteed by the equations of motion. The 4-vector $\chi^\mu$, sometimes called the particle dynamical velocity, is not in general co-linear 
with the particle kinematical velocity, $v^\mu$.

There is no equation of motion for $v^\mu$ and it should be determined indirectly from the other equations of motion. These equations result in 
the following relation~\cite{tod,ehl}
\begin{equation}\label{eq11}
v^\mu=(-v_\kappa\chi^\kappa)\left(\chi^\mu+\frac{2s^{\mu\nu}R_{\nu\rho\alpha\beta}
s^{\alpha\beta}}{4m^2+s^{\lambda\sigma}R_{\lambda\sigma\gamma\delta}s^{\gamma\delta}}
\chi^\rho\right)
\end{equation}
which simplifies further when a specific gauge, say
\begin{eqnarray}\label{e113}
v_\kappa\chi^\kappa=-1
\end{eqnarray}
is chosen. This is the gauge in which the instantaneous zero-momentum and the zero-velocity frames are simultaneous \cite{dix}. The MPD 
equations do not guarantee that $v^\mu v_\mu$ is constant.
\section{The energy-momentum tensor}
We construct a spinning fluid model in which the fluid equations of motion reduce to the MPD equations for a single spinning particle. We begin 
with a matter Lagrangian of the form
\begin{equation}
L=L(u^\mu,e_a^\mu,{\dot e}^\mu_a,X^i,\partial_\mu X^i)
\end{equation}
where $$u^\mu=\displaystyle\frac{dx^{\mu}}{dt}$$ are the fluid four-velocity, $X^i$ are the identity coordinates, and $e^\mu_a$ are the tetrad 
fields. The later satisfy
\begin{eqnarray}\label{e875}
e_\mu ^ae_{a\nu}&=&g_{\mu\nu},\nonumber\\e_{a\mu}e_b^\mu&=&\eta_{ab},
\end{eqnarray}
$a,b$ ranging from $0$ to $3$. The fields $X^i,i=1,2,3$ label the flow-lines in the sense that they identify the flow-line passing through any 
given point of space-time. To generate these, one can choose an arbitrary space-like hyper-surface equipped with a coordinate system $X^i$ and 
label each flow-line by the coordinate value of the point where it intersects that hyper-surface \cite{bro}.

We require
\begin{equation}\label{e853}
u^\mu\partial_\mu X^i=0
\end{equation}
which restricts the flow-lines to be directed along the integral curves of $X^i=const.$ The entropy per particle is assumed to be conserved 
along these curves.

The conjugate momentum is defined by
\begin{equation}
P_\mu=\frac{\partial L}{\partial u^\mu},\label{e88}
\end{equation}
and the spin (density) tensor by \cite{bai}
\begin{equation}
S_{\mu\nu}=e^a_\mu l_{a\nu}-e^a_\nu l_{a\mu}\label{e89}
\end{equation}
where $l^a_\mu=\displaystyle\frac{\partial L}{\partial{\dot e}^\mu_a}$.
This may be thought of as the density of the sum of individual particles spin $s^{\mu\nu}$. For later convenience we also define
\begin{equation}\label{x1}
\pi^\mu=\frac{P^\mu}{\sqrt{-P_\mu P^\mu}}
\end{equation}
representing a unit four-vector along the fluid momentum.

The action
\begin{equation}\label{e876a}
S=\frac{1}{\kappa}\int\sqrt{-g}Rd^4x+\int\sqrt{-g}Ld^4x
\end{equation}
describes the matter and the space-time dynamics. Requiring this action to be stationary under variation of tetrad field, and making use of the 
Einstein equation
\begin{equation}\label{equae}
G_{\mu\nu}=-\kappa T_{\mu\nu}
\end{equation}
we obtain
\begin{eqnarray*}
\sqrt{-g}T^{\mu\nu}e_{a\nu}\delta e^a_\mu=\delta_{e}(\sqrt{-g}L).
\end{eqnarray*}
By noting that the variation of $L$  with respect to the tetrad field gets contribution from both explicit dependence on the field and implicit 
dependence on it via covariant derivatives, we have
\begin{eqnarray*}
\delta_e(\sqrt{-g}L)=\sqrt{-g}\left(L\delta^\mu_\nu+\frac{\partial L}{\partial e^b_\mu}e^b_\nu+\frac{\partial L}{\partial{\dot e}^b_\mu}{\dot 
e}^b_\nu+\frac{1}{2}\nabla_\alpha {B^{\alpha\mu}}_\nu\right)e^\nu_a\delta e^a_\mu
\end{eqnarray*}
where $$B^{\alpha\mu\nu}=u^\mu S^{\nu\alpha}+u^\nu S^{\mu\alpha}+u^\alpha S^{\mu\nu}.$$
here the first term in the parentheses comes from variation of $\sqrt{-g}$, the second two terms from variational derivative with respect to 
tetrad and the last term from implicit tetrad dependence via connection coefficients, as is shown in the appendix. Terms like $u^\mu 
S^{\alpha\nu}$ are particular cases of the Belinfante-Rosenfeld tensor \cite{ray}.

On the other hand, by using the transformation properties of $L$, it can be shown that \cite{bai1}
\begin{eqnarray*}
\frac{\partial L}{\partial e^b_\mu}e_\nu^b+\frac{\partial L}{\partial{\dot e}^b_\mu}{\dot e}_\nu^b=\frac{\partial L}{\partial 
u^\nu}u^\mu-\frac{\partial L}{\partial(\partial_\mu X^i)}\partial_\nu X^i.
\end{eqnarray*}
Assume that the quantity $\partial_\mu X^i$ depends on the particle density number $n$. Thus $L$ depends on $n$ implicitly. So the last term in 
the above equation may be simplified further. Consider an infinitesimal flux tube $d^3X$ whose sections satisfy equation (\ref{e853}). Let 
$\gamma_{ij}$ denote the metric of this subspace and $\gamma$ its determinant. This metric can be obtained from the space-time metric by
\begin{equation}\label{e86}
\gamma^{ij}=g^{\mu\nu}\partial_\mu X^i\partial_\nu X^j.
\end{equation}
The number of particles in this tube is assumed conserved and is given by $N=n\sqrt{\gamma}$. We have
\begin{eqnarray*}
\frac{\partial L}{\partial(\partial_\mu X^i)}=\frac{\partial L}{\partial n}\frac{\partial n}{\partial\gamma}
\frac{\partial\gamma}{\partial(\partial_\mu X^i)}=n\frac{\partial L}{\partial n}\gamma_{ij}\partial^\mu X^j
\end{eqnarray*}
and hence
\begin{eqnarray*}
\frac{\partial L}{\partial(\partial_\mu X^i)}\partial_\nu X^i=n\frac{\partial L}{\partial n}\gamma_{ij}\partial^\mu X^j\partial_\nu X^i.
\end{eqnarray*}
We also have
\begin{equation}\label{e86a}
\gamma_{ij}\partial_\mu X^j\partial_\nu X^i=g_{\mu\nu}+u_\mu u_\nu
\end{equation}
note that $v^\mu$ is normal to the sections of the subspace. Thus we obtain
\begin{equation}\label{equa2}
\frac{\partial L}{\partial(\partial^\mu X^i)}\partial_\nu X^i=\frac{\partial L}{\partial n}nh^\mu_\nu
\end{equation}
where $h_{\mu\nu}=g_{\mu\nu}+u_\mu u_\nu$ is a projection tensor.

Putting the above results together, we get
\begin{eqnarray*}
\delta_e(\sqrt{-g}L)=\sqrt{-g}\left(\frac{\partial L}{\partial u^\nu}u_\mu
+L(g_{\mu\nu}+A_{\mu\nu})-n\frac{\partial L}{\partial n}h_{\mu\nu}+\frac{1}{2}\nabla_\alpha B^\alpha_{\mu\nu}\right)e^{a\nu}\delta e^\mu_a
\end{eqnarray*}
where we have added a vanishing term $$A_{\mu\nu}e^{a\nu}\delta e^\mu_a=0$$ to ensure that we will end up with a proper energy-momentum
tensor. The explicit form of $A_{\mu\nu}$ depends on the choice of tetrad. There are two natural choices for the tetrad field. These correspond 
to taking the time-like component of the tetrad along the directions of fluid four-velocity or four-momentum. (These choices have been used in
\cite{tab} for the case of spinning particles in a different context). We choose a tetrad for which $e^\mu_0$ coincides with the momentum 
direction, i.e., we set
\begin{equation}\label{e90}
e^\mu_0=\pi^\mu.
\end{equation}
In this case $A_{\mu\nu}e^{a\nu}\delta e^\mu_a=0$ is guaranteed by orthonormality of the tetrad components.
It turns out that this choice is consistent with the supplementary condition
\begin{equation}\label{equa1}
P_\mu S^{\mu\nu}=0.
\end{equation}
The other choice is given by $e^\mu_0=u^\mu$ which will result in an energy-momentum tensor as the one given in \cite{bai} together with a 
different supplementary condition.

Now let us define $f^\mu$ as the difference between the two velocities $u^\mu$ and $\pi^\mu$, 
\begin{equation}\label{tu}
f^\mu\equiv u^\mu-\pi^\mu. 
\end{equation}
This results in $$f_\mu f^\mu+2f_\mu\pi^\mu=0.$$  If we now neglect $f_\mu f^\mu$, an assumption based on the 
expectation that the effect of spin is not too strong, we get $$f_\mu\pi^\mu=0.$$ We now use this to obtain
\begin{eqnarray*}
u_\mu u_\nu e^{a\mu}\delta e^\nu_a&=&(\pi_\mu+f_\mu)(\pi_\nu+f_\nu)e^{a\mu}\delta e^\nu_a\\&=&\pi_\mu\pi_\nu e^{a\mu}\delta e^\nu_a
\end{eqnarray*}
We therefore obtain
\begin{eqnarray*}
T_{\mu\nu}e^{a\mu}\delta e^\nu_a=\left(P_\mu u_\nu+\left(L-n\frac{\partial L}{\partial n}\right)H_{\mu\nu}+\frac{1}
{2}\nabla_\alpha(u_\nu{S_\mu}^\alpha+u_\mu{S_\nu}^\alpha +u^\alpha S_{\nu\mu})\right)e^{a\mu}\delta e^\nu_a
\end{eqnarray*}
where in the right hand side we have interchanged $\mu$ and $\nu$, and have defined $$H_{\mu\nu}=g_{\mu\nu}+\pi_\mu\pi_\nu$$ (or, equivalently, 
$H_{\mu\nu}=e^i_\mu e_{i\nu}$). Here $P_\mu$ is given by relation (\ref{e88}). Also, the fluid pressure is defined by
\begin{equation}\label{e93}
P=L-n\frac{\partial L}{\partial n}.
\end{equation}
Thus we reach at the following energy-momentum tensor
\begin{equation}\label{e18}
T_{\mu\nu}=H_{\mu\nu}P+P_\mu u_\nu-\frac{1}{2}\nabla_\alpha({S^\alpha}_\mu u_\nu+{S^\alpha}_\nu u_\mu+S_{\mu\nu}u^\alpha)
\end{equation}
This is a generalization of the energy-momentum tensor given in \cite{bai1} for spinning dusts by taking the fluid pressure into account, and 
modifies the one given in \cite{obu} which rely on the Frenkel condition. It is also different from the energy-momentum tensor given in 
\cite{bai} by
the form of the projection tensor.

The second term in Eq.~(\ref{e18}) is the same as those of \cite{obu} and \cite{bai}. The third term is the divergence of some Belinfante-
Rosenfeld tensors. In the first term the usual perfect fluid term $g_{\mu\nu}+u_\mu u_\nu$ has been replaced by $g_{\mu\nu}+\pi_\mu\pi_\nu$.
\section{Equations of motion}
By requiring this energy-momentum tensor to be symmetric we obtain the equation of motion of the spin
\begin{equation}\label{eq1}
P_\mu u_\nu-P_\nu u_\mu=\nabla_\alpha(u^\alpha S_{\mu\nu})
\end{equation}
Demanding now the energy-momentum tensor to satisfy the conservation equation $\nabla^\nu T_{\mu\nu}=0$, yields
\begin{eqnarray}\label{eq2}
H_{\mu\nu}\nabla^\nu P+P\nabla^\nu(\pi_\mu\pi_\nu)+\nabla^\nu(P_\mu u_\nu)=-\frac{1}{2}{R}_{\mu\nu\alpha\beta}u^\nu S^{\alpha\beta}
\end{eqnarray}
This equation can also be derived by variation of world-lines.

By projecting Eq. (\ref{eq2}) parallel and normal to $\pi_\mu$, we can reduce it to the following equations
\begin{equation}
P\nabla_\mu\pi^\mu+\nabla_\mu(E u^\mu)=\frac{1}{2}R_{\mu\nu\alpha\beta}\pi^\mu u^\nu S^{\alpha\beta},\label{eq4a}
\end{equation}
\begin{eqnarray}
H_{\mu\nu}\nabla^\nu P+P\pi^\nu\nabla_\nu\pi_\mu +Eu^\nu\nabla_\nu\pi_\mu=-\frac{1}{2}H_{\mu\kappa}
{R^\kappa}_{\nu\alpha\beta}u^\nu S^{\alpha\beta}\label{eq4b}
\end{eqnarray}
respectively. Here $E$ denotes the norm of $P^\mu$. The first of these is a conservation equation and the second one is an equation of motion 
for the fluid momentum.

For a spinning dust, $P=0$ and $\rho=nm$, $m$ being the particles mass. Defining $$p^\mu=\frac{1}{n}P^\mu, s^{\mu\nu}=\frac{1}{n}S^{\mu\nu}$$ 
and using $\nabla_\mu(n u^\mu)=0$, equations (\ref{eq1}) and (\ref{eq2}) reduce to Eqs. (\ref{eq3a}) and~(\ref{eq3b}) respectively, the MPD 
equations.
\section{Applications}
In this section we apply the above formalism to a flat cosmological model. Consider a flat Friedman-Robertson-Walker space-time with the 
following line element
\begin{equation}\label{eq23}
ds^2=-dt^2+a^2(t)\delta_{ij}dx^idx^j.
\end{equation}
where $i,j$ run over the spacial dimensions, $a(t)$ is the scale factor. For co-moving matter we have $u^\mu=(1,0,0,0)$. Now as we do not expect 
the four-momentum to be co-linear with the four-velocity in general, we start with $P^\mu=(\rho(t),\epsilon p^i)$ where we will neglect 
$\epsilon^2$ and higher order terms. Note that this satisfies $u_\mu P^\mu=-\rho$, $\rho$ being the energy density. The later is the fluid 
analogue of (\ref{e113}). Thus we have $E=\rho$ and $$\pi^\mu=\left(1,\epsilon\frac{p^i}{\rho}\right).$$ By inserting this into the energy-
momentum tensor (\ref{e18}) we obtain
\begin{eqnarray}
T^{00}&=&\rho,\label{equat1}\\
T^{ii}&=&\frac{P}{a^2}\label{equat2}\\
T^{0i}&=&\frac{1}{2}\left(\epsilon(2w+1)p^i-\partial_t S^{0i}-\frac{2\dot a}{a}S^{0i}\right)\label{equat3}
\end{eqnarray}
where $w$ is the so-called equation of state parameter given by $w=\frac{P}{\rho}$. Now the Einstein's equation gives
\begin{eqnarray}
\rho(t)&=&3\kappa^{-1}\left(\frac{\dot a(t)}{a(t)}\right)^2,\label{el1a}\\
P(t)&=&-\kappa^{-1}\left(\frac{\dot a^2(t)}{a^2(t)}+2\frac{\ddot a(t)}{a(t)}\right),\label{el2a}
\end{eqnarray}
and
\begin{eqnarray}
\epsilon(2w+1)p^i(t)=\partial_t S^{0i}+2\frac{{\dot a}(t)}{a(t)}S^{0i}\label{el3a}.
\end{eqnarray}
On the other hand, the spin equation of motion (\ref{eq1}) yields
\begin{eqnarray}
\frac{dS^{ij}}{dt}+5\frac{\dot a}{a}S^{ij}&=&0,\label{equat5}\\
\frac{dS^{0i}}{dt}+4\frac{\dot a}{a}S^{0i}&=&-\epsilon p^i.\label{equat6}
\end{eqnarray}
The first of these immediately results in
\begin{equation}\label{hk1}
S^{ij}(t)=\sigma^{ij}(a(t))^{-5}
\end{equation}
in which $\sigma^{ij}$ are integration constants. The second equation when combined with equation (\ref{el3a}) yields
\begin{eqnarray}
S^{0i}&=&\epsilon\sigma^{0i}(a(t))^{-n}\label{ll2}\\
p^i&=&-\frac{1}{w+1}\sigma^{0i}{\dot a(t)}(a(t))^{-(n+1)}\label{ll3}
\end{eqnarray}
where $$n=\frac{4w+3}{w+1}.$$
From the translational equation of motion (\ref{eq4b}) we obtain
\begin{equation}
\epsilon(1+w)\frac{dp^i}{dt}+\epsilon(4+5w)\frac{\dot a}{a}p^i=-\frac{\ddot a}{a}S^{0i}\label{ll1}
\end{equation}
for which we have already a solution, relations (\ref{ll2}) and (\ref{ll3}). Inserting this solution back into the differential equation shows 
that it is valid for any $w\neq-1$. One can easily check that the above solutions of the equations of motion satisfy the conservation equation 
(\ref{eq4a}) however they are consistent with the supplementary condition (\ref{equa1}) only for 
\begin{equation}\label{d45}
\sigma^{0i}=0
\end{equation}
i.e. we end up with a four-momentum parallel to the four-velocity in this case. This is due the maximal symmetries of the space-time described 
by (\ref{eq23}). From equation (\ref{hk1}) we have $S^2=\frac{1}{2}S_{\mu\nu}S^{\mu\nu}=\frac{1}{2}\sigma_{ij}\sigma^{ij}(a(t))^{-6}$. Thus
the fluid spin (the spin density integrated over a spacial volume) is space-like and goes like $(a(t))^{-3}$. 

For both the above model and the usual perfect-fluid model the application of the Einstein's equation gives the same expression for the energy 
density and pressure in terms of the scale factor. Thus, in the present context the evolution of $a(t)$ is the same as the standard model with a 
usual perfect fluid. This is in agreement with the result obtained in \cite{obu}. However it would be possible to obtain a different evolution
for the scale factor by applying an averaging on spins scheme, as it has been done in \cite{gasper} for the case of Einstein-Cartan theory.  
\section{Discussion}
We have constructed a consistent formulation of spinning fluids compatible with the Tulczyjew-type supplementary condition by starting  
with a rather general Lagrangian depending on a tetrad field that its gyration simulates the spin. There are two natural choices for 
the time-like component of this tetrad. These correspond to the velocity and the momentum directions and result in different supplementary 
equations. Our choice advocate the supplementary equation $P_\mu S^{\mu\nu}=0$. The resulting energy-momentum tensor reduces to the usual 
perfect fluid energy-momentum tensor in the case where the spin is turned off and to the energy-momentum tensor introduced in Ref.\cite{bai1} 
for spinning dusts, and the relevant equations of motion reduce to the MPD equations. The general form of this energy-momentum tensor and 
consequent equations of motion differ from those of Ref. \cite{bai} mainly due to different forms of the projection tensors.
We have shown that the equations of motion admit a flat FRW space-time as a solution. For this solution the four-velocity and four-momentum are 
co-linear due to maximal symmetries of the solution. The dynamics of space-time in this model is similar to the one obtained by a usual perfect 
fluid model but applying an averaging process could result in different evolutions. It would be interesting to find solutions for which the 
the fluid velocity and momentum are not co-linear. Research is underway along this direction. 

\section*{Acknowledgements}
I would like to thank an anonymous referee of General Relativity and Gravitation for valuable comments.
\appendix
\section{Tetrad variation of the Lagrangian}
In this appendix we calculate the variation of $L$ with respect to tetrad. We have
\begin{eqnarray*}
\delta_{e} L=\frac{\delta L}{\delta e_\mu^a}\delta e_\mu^a-\frac{\partial L}
{\partial{\dot e}^a_\mu}\,u^\alpha e^{a\nu}\delta\Gamma_{\nu\alpha\mu}.
\end{eqnarray*}
But
\begin{eqnarray*}
\frac{\delta L}{\delta e_a^\mu}\,\delta e_a^\mu&=&\left(\frac{\partial L}{\partial
e_\mu^a}-\nabla_\alpha\left(u^\alpha\frac{\partial L}{\partial{\dot e}_\mu^a}\right)\right)
\delta e_\mu^a\\&=& \left(\frac{\partial L}{\partial e_\mu^b}\delta^b_a-
\delta^b_a\nabla_\alpha\left(u^\alpha\frac{\partial L}{\partial{\dot e}_\mu^b}\right)\right)
\delta e_\mu^a\\&=&\left(\frac{\partial L}{\partial e_\mu^b}e^b_\nu e_{a\nu}-e^b_\nu
e_{a\nu}\nabla_\alpha\left(u^\alpha\frac{\partial L}{\partial{\dot e}_\mu^b}\right)\right)
\delta e_\mu^a\\&=& \left(\frac{\partial L}{\partial e_\mu^b}e^b_\nu +\frac{\partial L}
{\partial{\dot e}^b_\mu}{\dot e}^b_\nu-\nabla_\alpha\left(u^\alpha e^b_\nu\frac{\partial L}
{\partial{\dot e}_\mu^b}\right)\right)e_a^\nu\delta e_\mu^a
\end{eqnarray*}
and
\begin{eqnarray*}
u^\alpha e^{a\nu}\frac{\partial L}{\partial{\dot e}^a_\mu}\delta\Gamma_{\nu\alpha\mu}
&=&\frac{1}{2}u^\alpha e^{a\nu}\frac{\partial L}{\partial{\dot e}^a_\mu}
\delta(\nabla_\mu g_{\alpha\nu}+\nabla_\alpha g_{\nu\mu}+\nabla_\nu g_{\alpha\mu})\\&=&
-\frac{1}{2}\nabla_\alpha\left(u^\mu e^{a\nu}\frac{\partial L}{\partial{\dot
e}^a_\alpha}+u^\alpha e^{a\nu}\frac{\partial L}{\partial{\dot e}^a_\mu}
-u^\nu e^{a\alpha}\frac{\partial L}{\partial{\dot e}^a_\mu}\right)
\delta g_{\mu\nu}\\&=&-\frac{1}{2}\nabla_\alpha\left(u^\mu e^{a\nu}
\frac{\partial L}{\partial{\dot e}^a_\alpha}+u^\nu e^{a\mu}\frac{\partial
L}{\partial{\dot e}^a_\alpha}+u^\alpha e^{a\nu}\frac{\partial L}
{\partial{\dot e}^a_\mu}\right.\\&&\left.+u^\alpha e^{a\mu}\frac{\partial L}{\partial{\dot e}^a_\nu}
-u^\nu e^{a\alpha}\frac{\partial L}{\partial{\dot e}^a_\mu}-u^\mu e^{a\alpha}
\frac{\partial L}{\partial{\dot e}^a_\nu}\right)e_{a\nu}\delta e^a_\mu\\&=&
-\frac{1}{2}\nabla_\alpha\left(u^\alpha e^{a\mu}\frac{\partial L}{\partial{\dot
e}^a_\nu}+u^\mu S^{\nu\alpha}+u^\alpha e^{a\nu}\frac{\partial L}{\partial{\dot
e}^a_\mu}+u^\nu S^{\mu\alpha}\right)e_{a\nu}\delta e^a_\mu
\end{eqnarray*}
where we have omitted a surface term. Thus
\begin{eqnarray*}
\delta_e L=\left(\frac{\partial L}{\partial e^b_\mu}e^{b\nu}+
\frac{\partial L}{\partial{\dot e}^b_\mu}{\dot e}^{b\nu}+\frac{1}{2}\nabla_\alpha
(u^\mu S^{\nu\alpha}+u^\nu S^{\mu\alpha}+u^\alpha S^{\mu\nu})\right)e_{a\nu}\delta e^a_\mu.
\end{eqnarray*}


\begin{thebibliography}{10}
\bibitem{dix} W.G. Dixon, Proc. R. Soc. London, ser. A 314 (1970) 499.
\bibitem{ray} J.R. Ray, L.L. Smalley, Phys. Rev. D 26 (1982) 2619.
\bibitem{sma} J.R. Ray, L.L. Smalley, Phys. Rev. D 27 (1983) 1383.
\bibitem{obu} Y.N. Obukhov, O.B. Piskareva, Class. Quantum Grav. 6 (1989) L15.
\bibitem{bai} I. Bailey, Ann. Phys. 119 (1979) 76.
\bibitem{kop}W. Kopczynski, Phys. Rev. D 34 (1986) 352.
\bibitem{tod} K.P. Tod, F. de Felice, M. Calvani, Nuovo Cimento, 34 B (1976) 365.
\bibitem{ehl}J. Ehlers, E. Rudolph, Gen. Rel. Grav. 8 (1977) 197.
\bibitem{bro} J.D. Brown, Class. Quantum Grav. 10 (1993) 1579.
\bibitem{bai1} I. Bailey, W. Israel, Commun. math. Phys. 42 (1975) 65.
\bibitem{tab}G.E. Tauber, Int. J. Theo. Phys. 27 (1988) 335.
\bibitem{gasper} M. Gasperini, Phys. Rev. Lett. 56 (1986) 2873.
\end{thebibliography}
\end{document}